\documentclass[letter]{aa} 
\usepackage{amsmath}
\newcommand{\h}{$h^{-1}$}

\usepackage{booktabs} 
\usepackage{subcaption} 

\usepackage{graphicx}
\usepackage{txfonts}
\begin{document}

   \title{A local Universe catalogue of structures and voids dynamically identified using Cosmic-Flows4++ZOA peculiar velocities}

   \author{ A.M. Hollinger\inst{1,2}\thanks{Amber.Hollinger@anu.edu.au}, H. M. Courtois\inst{1}}

   \institute{Universit\'e Claude Bernard Lyon 1, IUF, IP2I Lyon, 4 rue Enrico Fermi, 69622 Villeurbanne, France\\
    \and Research School of Astronomy and Astrophysics, Australian National University, Canberra, ACT 2611, Australia\\
             }

   \date{Received XXX; accepted XXX}
\titlerunning {Local Universe voids and structures}
\authorrunning{Hollinger \& Courtois}
 
  \abstract
 {

Cosmic voids and superclusters are among the largest structures in the Universe and provide complementary probes of the growth of large-scale structure and the underlying gravitational field. We present a comprehensive catalogue of the main local Universe cosmic web voids and knots, using the updated CosmicFlows-4++ Zone of Avoidance (CF4++ZOA) catalogue out to redshift z=0.1. To do this we use the V-web algorithm which provides a dynamically motivated map of the the local cosmic web, tracing the major expanding and converging regions of the local Universe. The robustness of the catalogue is assessed using the ensemble of Hamiltonian Monte Carlo realizations of the CF4++ZOA reconstruction. We additionally remove any structure that exceeds the survey boundaries and present catalogues of 37 voids and 42 knot regions within the reconstructed survey volume.  The identified emptying regions (voids) have effective radii ranging from 13 to 38 \h Mpc. The high density converging regions (knots) have volumes ranging from 10$^4$ to 3.3$\times 10^5$ $h^{-3}$ Mpc$^3$.

}
 


   \maketitle
\section{Introduction}

On the largest scales, matter in the observable universe forms a complex, web-like pattern, with filaments linking dense galaxy clusters \citep{bond_how_1996}. Between these grand structures lie vast, almost empty regions where surveys detect relatively few galaxies \citep{joeveer_spatial_1978, gregory_comaa1367_1978,kirshner_million_1981}. These cosmic voids, which occupy most of the Universe’s volume \citep{de_lapparent_slice_1986,geller_mapping_1989}, arise from gravitational instability in primordial under-dense regions, in a way analogous to how clusters form from the collapse of overburden regions \citep{zeldovich_gravitational_1970,van_de_weygaert_cosmic_2011}.  Together, these structures trace the growth of primordial density fluctuations through gravitational instability and provide a powerful probe of structure formation and cosmology \citep{zeldovich_gravitational_1970, van_de_weygaert_cosmic_2011, pisani_cosmic_2019}.

Mapping the intricacies of the cosmic web is a central aim of observational cosmology, driving deep and wide-area sky surveys \citep[e.g.,][]{the_dark_energy_survey_collaboration_dark_2005, lsst_science_collaboration_lsst_2009, levi_desi_2013, amendola_cosmology_2013} as well as a variety of numerical simulations \citep{cai_towards_2009, potter_pkdgrav3_2017,takahashi_full-sky_2017,racz_complementary_2023,schaye_flamingo_2023}. Among the largest and most extreme features of this large-scale structure are the superclusters of galaxies and cosmic voids. 

Cosmic voids have attracted growing interest as cosmological probes in recent years due to their sensitivity to several fundamental phenomena, including the Alcock--Paczyński effect \citep{lavaux_precision_2012, sutter_first_2012,sutter_measurement_2014, hamaus_constraints_2016,mao_cosmic_2017,nadathur_beyond_2019}, dark energy \citep{pisani_counting_2015,verza_void_2019}, baryon acoustic oscillations \citep{lavaux_precision_2012,sutter_first_2012,melchior_first_2014,pisani_counting_2015,hamaus_constraints_2016,zhao_completed_2022}, and weak lensing  \citep{melchior_first_2014,chantavat_void_2017}. They also provide a unique environment for galaxy evolution, with galaxies in voids exhibiting distinct properties from those in denser regions \citep{hoyle_luminosity_2005,rojas_spectroscopic_2005,habouzit_properties_2020}, and their interior dynamics remain in the linear regime for a relatively long period \citep{goldberg_simulating_2004}.
Voids are commonly identified using geometric methods, including empty-sphere searches \citep{el-ad_voids_1997,hoyle_voids_2002} and watershed algorithms based on Voronoi tessellations \citep{neyrinck_zobov_2008,sutter_vide_2015}. While these approaches define voids geometrically, simulations enable dynamically motivated definitions based on gravitational collapse \citep{sheth_hierarchy_2004}. 

The over-dense counterparts to voids are galaxy superclusters, the largest coherent structures in the cosmic web, which offer vital insights into non-linear structure formation and the transition from filament- to cluster-dominated environments \citep{bahcall_large-scale_1988,einasto_supercluster-void_1997,liivamagi_sdss_2012}. Superclusters trace the nodes and intersections of the cosmic web, marking regions where matter assembly and environmental effects are most pronounced, providing a complementary perspective on large-scale gravitational evolution and the growth of structure \citep{bond_how_1996,einasto_multimodality_2012}. They are typically identified using friends-of-friends algorithms \citep[e.g][]{zeldovich_giant_1982, einasto_superclusters_2007, bohringer_classix_2022}, density-threshold methods \citep{einasto_richest_2007,liivamagi_sdss_2012}, or connectivity-based approaches \citep[e.g.][]{barrow_minimal_1985,naidoo_beyond_2020}. Dynamical classifications based on the velocity shear tensor offer an alternative framework by linking structures directly to anisotropic gravitational collapse \citep{hahn_properties_2007, hoffman_kinematic_2012,pomarede_cosmic_2017}. In this work, we use such a dynamical approach to identify both under-dense and over-dense regions, enabling a unified study of the largest structures that define the cosmic web.

We characterize the large-scale structure of the reconstructed density field using the velocity cosmic web (V-web) formalism introduced by \cite{zaroubi_wiener_1999} and the methodology described in \cite{courtois_sociology_2023,  courtois_search_2025}. The V-web is based on the eigenvalue analysis of the velocity shear tensor derived from the three-dimensional peculiar velocity field. This approach classifies the cosmic web into voids, sheets, filaments, and knots in a fully dynamical framework, directly linked to the underlying gravitational flow. In this framework, the local cosmic web environment is determined by the signs and relative amplitudes of the three eigenvalues of the velocity shear tensor. Regions where all three eigenvalues are positive correspond to expanding flows and are classified as voids, while regions where all three eigenvalues are negative correspond to converging flows and define knots. Intermediate configurations, with two positive and one negative eigenvalue or vice versa, are identified as sheets and filaments, respectively.

This letter is structured as follows: Section 2 describes the data and the methodology used in this study. Section 3 presents the results and offers a discussion of our findings. Finally, Section 4 provides a summary of our main conclusions.

\section{Data and Methodology}\label{sec:data}

To recover the large-scale density and velocity fields from galaxy peculiar velocity surveys, we use  a purely linear iterative  Hamiltonian-Monte-Carlo (HMC) Bayesian reconstruction method that is presented in \citep{graziani_peculiar_2019,courtois_gravity_2023}. The technique uses galaxy distances gathered in the updated CosmicFlows-4 Zone of Avoidance (CF4++ZOA) catalogue \citep{hollinger_hidden_2026}, to solve the linearized continuity and Poisson equations. The method additionally assumes that both galaxy bias and structure growth are linear.  To achieve a stable convergence on the free parameters, more than 10,000 HMC steps were computed to obtain an overall mean density and peculiar velocity fields. 
For this work, we use both the individual HMC and overall mean fields of CF4++ZOA. These have a grid resolution of $128^3$ over a length of 1000 \h Mpc, resulting in a voxel length of 7.8 \h Mpc. 

\begin{figure}
    \centering
    \includegraphics[width=1\linewidth]{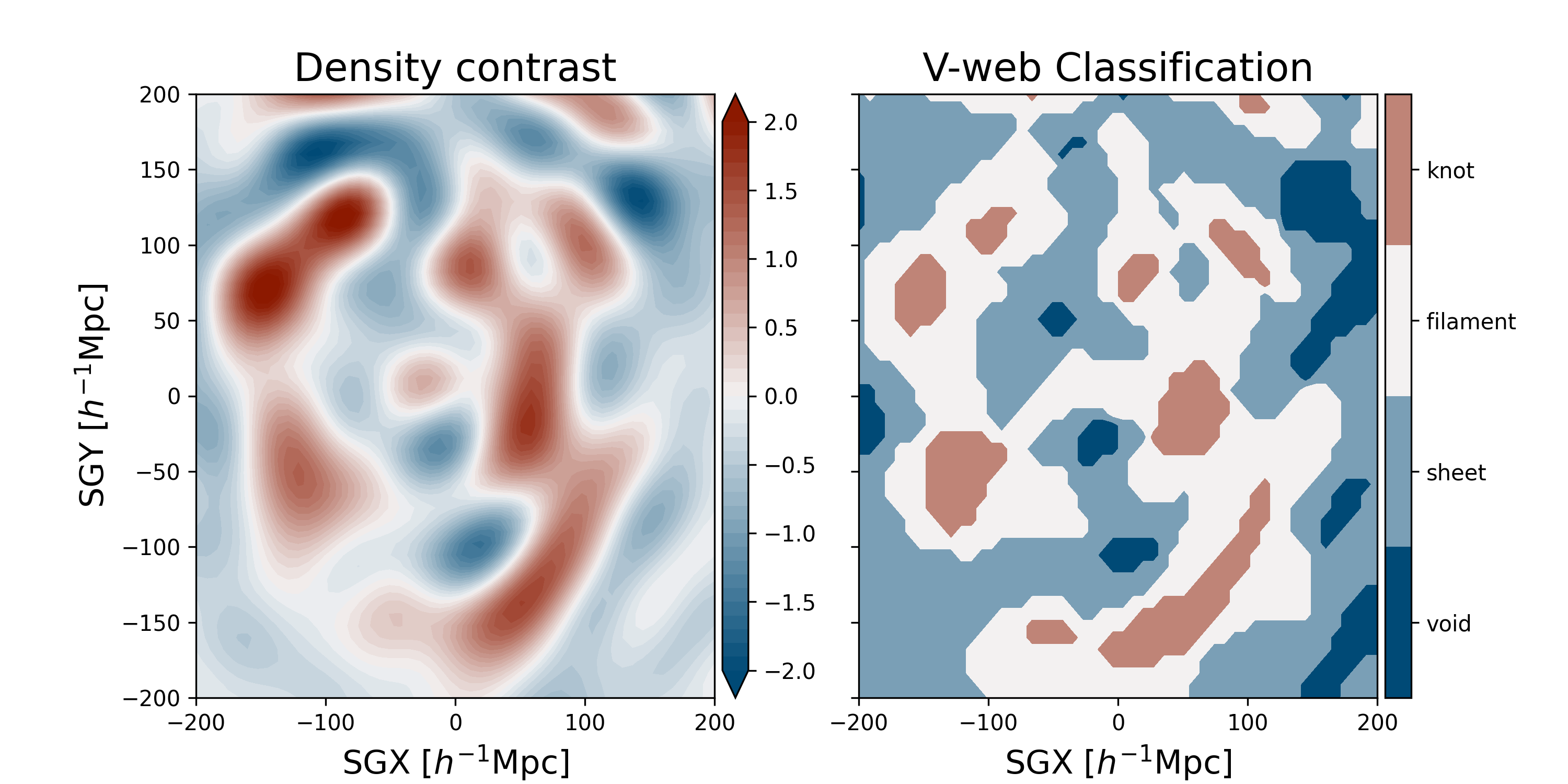}
    \caption{Comparison of the CF4++ZOA local density perturbation field (left) and the V-web classification (right), through the SGZ=-27 \h Mpc plane.}
    \label{fig:d_vs_vweb}
\end{figure}

\subsection{Classifying the Cosmic Web}

The V-web algorithm developed by \citep{hoffman_kinematic_2012,hoffman_dipole_2017} is a technique used to classify the large-scale structure of the cosmic web based on the reconstructed velocity field. Originally introduced within the context of cosmological structure formation it encapsulates the local tidal influences resulting from matter flows, and identifies regions as voids, sheets, filaments, or nodes by analysing the eigenvalues of the velocity shear tensor:
\begin{equation}
    \Sigma_{ij} = -\frac{1}{2H_0}\left(\frac{\partial v_i}{\partial r_j} + \frac{\partial v_j}{\partial r_i} \right) 
\end{equation}
where partial derivatives of velocity ($v$) are calculated along the ith and jth directions of the orthogonal supergalactic Cartesian axes. The Hubble parameter $H_0 = 100 h$ Mpc scales the tensor at each redshift and renders it dimensionless, while the negative sign incorporated ensures that postive eigenvalues are associated with collapse.

The eigenvectors of the shear, with associated eigenvalues ordered such that $\lambda_1 > \lambda_2 > \lambda_3$, define the principal axes of the collapse and expansion. 
These three eigenvalues, are compared against a threshold ($\lambda_{th}$) to determine the environment. The value of which is a free parameter that has typically between 0 and 1 and chosen so that the V-web produces a visual reproduction of the expected structures. Given that $\lambda_{th} = 0$ is a conventional choice in the linear regime \citep{garcia-bellido_information_2026}, is consistent with previous CosmicFlows V-web analyses and successfully reproduces the anticipated large-scale structures, it is the value is adopted in this work. A discussion of alternative threshold choices is provided in Appendix A.
This classification directly relates to the dynamical state of the region, if $\lambda_{th} \leq \lambda_{i}$ in a given direction, it indicates expansion along that axis; conversely if $\lambda_{th} \geq \lambda_{i}$ it indicates compression. When expansion occurs in all three dimensions, it points to the presence of large voids, whereas compression across all three axes corresponds to densely packed knots. 
This approach provides a robust, physically motivated way to distinguish structures without relying solely on density-based criteria. Figure \ref{fig:d_vs_vweb} compares the reconstructed CF4++ZOA local density field and its V-web classification, illustrating how the velocity shear tensor decomposition captures the main structures of the cosmic web and broadly follows the density distribution.




\subsection{Methodology}\label{sec:methodology}

\begin{figure}
    \centering
    \includegraphics[width=.8\linewidth,trim={0cm 0.5cm 0.5cm 0cm}]{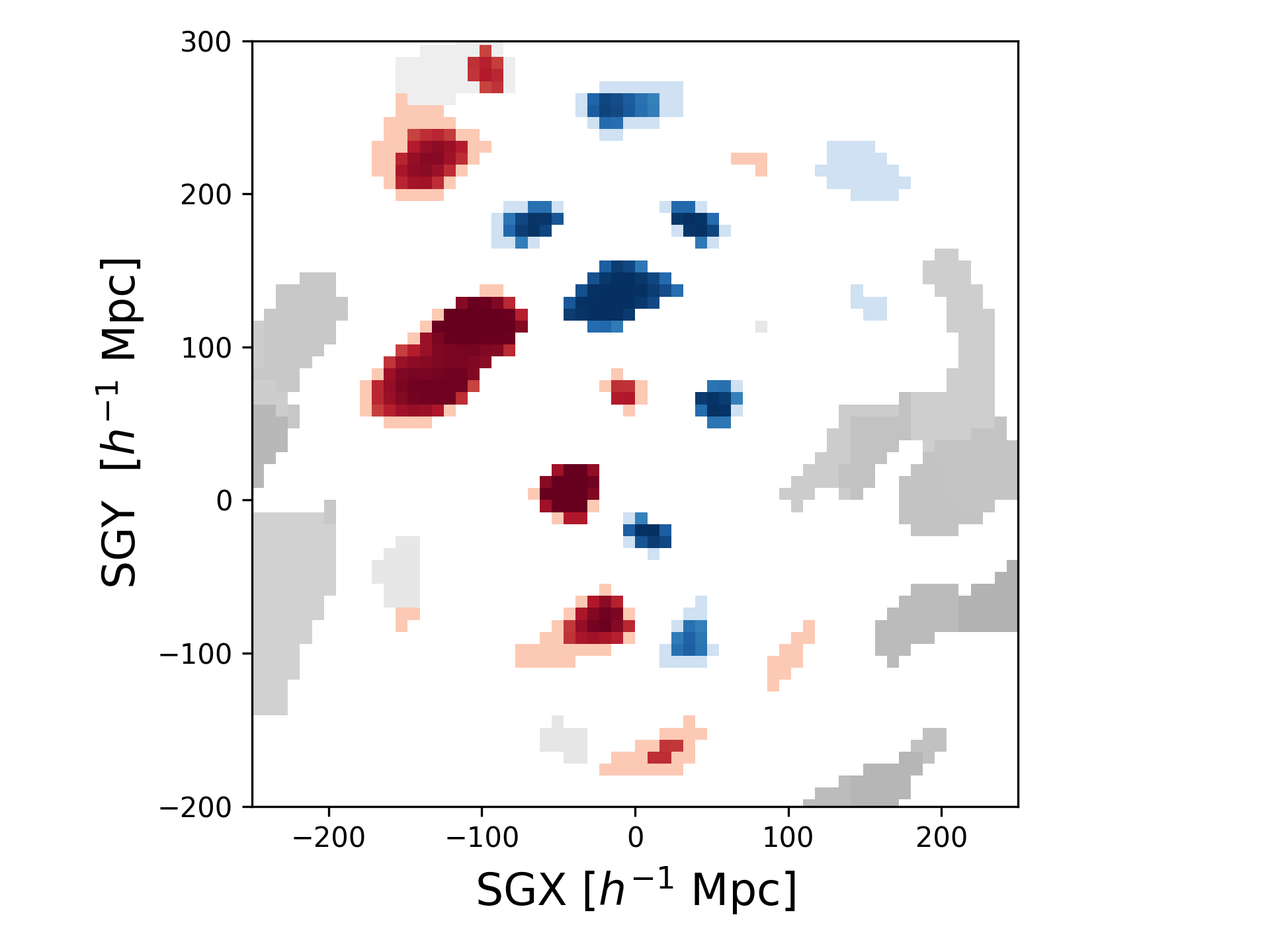}
    \caption{The shaded regions depict the spatial distribution of voids and knots identified in the full CF4++ZOA reconstruction assuming $\lambda_{\rm th}=0$, centred on SGZ=4 \h Mpc.  Light-coloured blue (voids) and red (knot) regions indicate structures that satisfy the selection criteria described in Section~\ref{sec:methodology}. The corresponding dark-coloured regions denote regions which were classified as the corresponding structure type in at least 68\% of the HMC realizations.}

    \label{fig:comp_50}
\end{figure}

Our procedure for determining the location and size of the void and knot regions can be summarized as follows: 
\vspace{-1.5mm}
\begin{itemize}
    \item 
   We systematically compute the eigenvalues of the velocity shear tensor and compute each voxels V-web classification for  a sequence of $\lambda_{th}$ parameters, starting from the lowest 
   value from which a void 
   region is detected and incrementally increasing up to a maximum 
   value of 0. For each  $\lambda_{th}$ value, we evaluate the eigenvalues at every point within the spatial domain, generating a comprehensive set of data that captures how the local properties of the field evolve with the variation of the parameter.
   \item
   For each voxel we determine the minimum $\lambda_{th}$ a region would need in order to be classified as a void, up to a maximum of 0.  Consequently, the local minima in this hierarchical field correspond to the regions exhibiting the greatest expansion in the velocity field and are adopted as candidate void centres. 
    \item
   To ensure meaningful separation of distinct voids, we enforce a minimum spatial distance between these minima, of 20 \h Mpc. ensuring that each minimum corresponds to a physically distinct void region.

    \item
   Starting from each identified minimum, we perform a flood fill algorithm of the grid. In each iteration, 
   we expand the assigned regions outward by one layer of neighbouring points, effectively ``filling'' the voids from their seed points. We assign unique group labels to each region as it grows, ensuring that the voids are distinctly identified and separated. This simultaneous, level-by-level expansion helps to accurately map the full extent of each void in the survey volume.

    \item
    Finally, we ensure that we are not including void regions that are poorly identified due to the boundaries of the CF4++ZOA survey.   Figure \ref{fig:comp_50} shows the regions where voids (blue) and knots (red) are identified in more than 68\% (representing a 1$\sigma$ confidence level) of our HMC realizations. In contrast, the void (dark) and knot (light) regions near the survey boundaries are  spurious features, and do not appear in the majority of the reconstructions, and hence need to be excluded from our final catalogues.  In order to remove these non-robust structures, we calculate the V-web for each HMC realization and compare the resulting classifications across the ensemble.  If a void is recovered in fewer than 68\% of the HMC realizations, or if more than 50\% of its volume lies beyond the survey boundaries, as determined by sampling the distribution of the CF4++ZOA galaxies, it is excluded from the final catalogue. 


\end{itemize}
We repeat this procedure for the knot regions by: searching for local maxima, starting from the highest value of $\lambda_{th}$ and decreasing to the same minimum threshold of 0. This process results in the identification of 37 distinct void regions and 42 knot regions. The full catalogues of which are presented in Appendix \ref{app:A}, while sample summary tables are included in the main text.

\begin{table*}[htbp]
    \centering
    \caption{Extract of void (a) and knot (b) properties from the full catalogues.}
    \label{tab:voids}
    \begin{subtable}[c]{\textwidth}
    \centering
    \begin{tabular}{cccccccc}
    \hline
    Void ID &  Name & R.A  & Dec. & $D_c$  & SGX,SGY,SGZ & Volume & Radius \\
    
    \hline
    10 & Local Void & 296.7 & -1.1 & 93  & (20,-12,90) & 4.91  $\times10^4$ & 22.7\\
    13 & Northern Local Void & 328.4 & -47.7 & 34  & (-4,-27,20) & 3.53$\times10^4$  & 20.3\\
    28 & Sculptor & 9.9 & -68.5 & 104  & (35,-98,4) & 3.15$\times10^4$  & 19.6 \\
    \hline
    \end{tabular}
    \caption{Extract of void properties from Table \ref{tab:vv}: (1) ID;
    (2) Common Name;
    (3-5) Equatorial coordinates [$^\circ$] and distance  [\h Mpc] of local extrema; (6) Supergalactic Cartesian coordinates [\h Mpc]; 
    (7) Total volume  [ $h^{-3}$Mpc$^3$];
    (8) Effective radius  [\h Mpc].}\label{sub:void}
        \end{subtable}
        \hfill 
        \begin{subtable}[c]{\textwidth}
       \centering

    \begin{tabular}{cccccccc}
    \hline
    Knot ID &  Name & R.A  & Dec. & $D_c$  & SGX,SGY,SGZ & Volume & Mass \\
    
    \hline
    2 & Shapley  & 199.8 & 46.9 & 155 & (-105,113,4) & 17.2& 5.90$\times10^{16}$  \\
    3 & Vela  & 123.1 & -8.4 & 201 & (-121,-12,-160) & 14.5 & 4.42$\times10^{16}$ \\
    4 & Vela$^*$   & 153.7 & 13.9 & 186 & (-129,59,-121) & 14.3 & 4.25$\times10^{16}$  \\
    \hline
    \end{tabular}
    \caption{Extract of knot properties from Table \ref{tab:structures}: Columns 1-7 same as (a);
    (8) Enclosed mass [M$_{\odot}$]. }\label{sub:knot}

    \end{subtable}

\end{table*}

\section{Results}

The large-scale structure inferred from the shear of the velocity field, the cosmic V-web, provides a simplified but physically motivated description of the matter distribution in the Universe. Although it does not capture the full level of detail visible in redshift maps of galaxy distributions  or in high-resolution $N$-body simulations, it nevertheless retains the dominant dynamical features of the cosmic web. Fine substructure, small filaments, and irregular boundaries are often smoothed out in the reconstructed velocity field, particularly when the available peculiar velocity data are sparse or noisy. Our effective spatial resolution of 7.9 \h Mpc limits our sensitivity to smaller structures and the boundaries of the identified knots and voids.   Additionally, given that the CF4++ZOA dataset is a compilation of various surveys there are regions which have few to no galaxies, limiting the reconstruction of the local velocity and density fields. While the enforcement of survey limits and the use of HMC realizations to discard non-robustly defined structures mitigates these effects, future reconstructions at higher resolution using larger and deeper velocity samples will allow a more complete characterization of the cosmic web.
However, because peculiar velocities directly correlate to the underlying gravitational potential, the V-web remains highly effective at tracing coherent large-scale flows which govern structure formation. It is capable of robustly identifying the major attractors and repellers associated with dense knots and deep voids. These flow structures are often less apparent in traditional redshift surveys, where peculiar velocity distortions and projection effects can obscure the true dynamical connectivity of large-scale structure.


Table \ref{tab:voids} lists the centres of the void (a) and knot (b) regions, as determined from the extrema of the hierarchical field, in equatorial sky coordinates and comoving distance, as well as in supergalactic Cartesian coordinates. The total volume of each region is also provided. 
Table \ref{sub:void} includes the common names of the voids, as determined from the literature, along with their effective radii calculated from their volumes. Table \ref{sub:knot} instead lists the corresponding supercluster region of \cite{hollinger_hidden_2026}, hereafter H26, and mass enclosed within the given volume.
For the purposes of this letter, knots connected within high-density regions containing distinct local maxima are treated as independent and separate. Structures that were combined into the primary over-density in that work are marked with an asterisk. 

We focus our findings of the knot regions to the key structures of H26. We find that the primary collapsing region in the Shapley supercluster region of $6\times10^{16}$ M$_\odot$   to be comparable to the $5\times10^{16}$ M$_\odot$ found by \citep{proust_structure_2006}. Including the two substructures we recover a mass similar to that within the isocontour $\delta > 1.5$ of H26. We additionally find comparable masses to the X-ray luminosity estimates of \citep{bohringer_classix_2022} of Hercules ($4\times10^{16}$ M$_\odot$).  Overall, the characteristic masses reported here are slightly higher than those reviewed by \cite{einasto_galaxy_2025}, likely reflecting methodological differences, as most supercluster mass estimates are based on luminosity rather than peculiar velocity or density fields.
The choice of $\lambda_{th}$ also affects our sizes and masses if we adopted $\lambda_{th} = 0.44$ as was done by \cite{hoffman_kinematic_2012}, Shapley's core would lose 30\% of its volume and 23\% of its mass.

\begin{figure}
    \centering
    \includegraphics[width=1\linewidth,trim={1.5cm 2cm 1cm 2cm}]{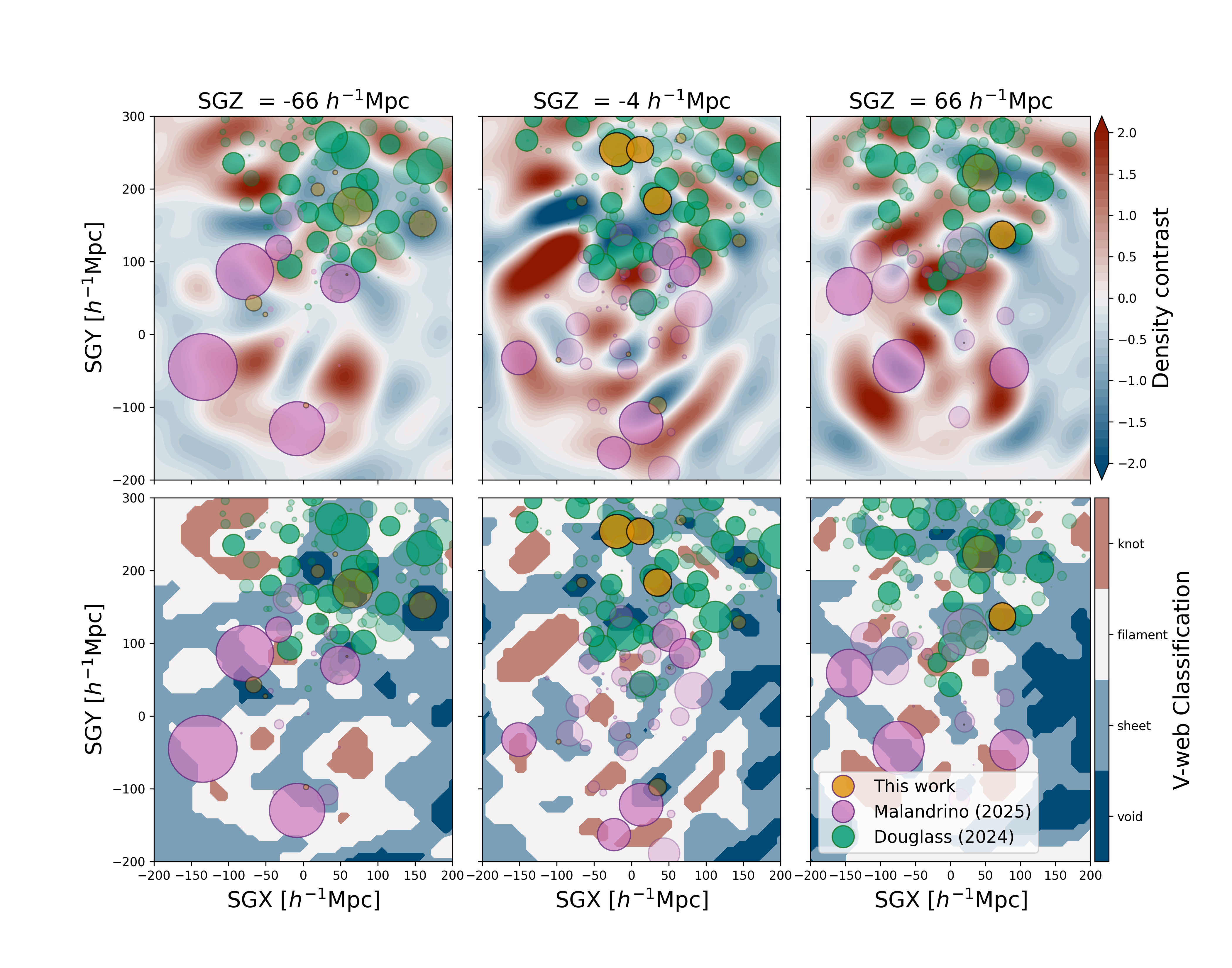}
    \caption{Comparison of the void locations identified in this work (orange) with those reported by \cite{malandrino_bayesian_2026} (pink) and \cite{douglass_updated_2023} (green), centred on SGZ=-66 \h Mpc (left), 4 \h Mpc (middle), and 66 \h Mpc (right). The Darker-coloured spheres represent voids whose centres lie within the 7.8 \h Mpc-thick slice, while lighter-coloured spheres show voids located above or below the slice whose volumes intersect the displayed plane. }
    \label{fig:2mppvoids}
\end{figure}

Figure \ref{fig:2mppvoids} compares the void locations identified in this work (orange) to those reported by \cite{malandrino_bayesian_2026},(hereafter M26), (pink) and \cite{douglass_updated_2023}(hereafter D23) (green) through various slices of the SGZ plane. The latter catalogues consist of 100 and 535 distinct voids, respectively, and identify voids using VIDE \citep{sutter_vide_2015}, a watershed algorithm void-finder which implements an enhanced version of ZOBOV \citep{neyrinck_zobov_2008}. 
The darker spheres denote voids whose centres lie within the same sliced region of our reconstruction, while the lighter spheres represent voids located above or below the plane that nevertheless intersect the displayed slice. 
Despite significant overlap with the latter two catalogues, agreement is limited in both the number and sizes of identified voids, reflecting the fundamentally different methodologies used.
Our analysis  traces coherent gravitational flows and regions of expansion in the CF4++ZOA reconstruction. In contrast, M26 identify voids from a Bayesian reconstruction of the density field using 2M++ data \citep{carrick_cosmological_2015}. As a result their catalogue is more directly sensitive to local density minima and the adopted statistical priors. D23 use a volume-limited subsample of the Sloan Digital Sky Survey Data Release 7 \citep{abazajian_seventh_2009} and identifies voids with their Void Analysis Software Toolkit \citep[VAST]{douglass_vast_2022}.  While both M26 and D23 use similar void finding methods the sizes of the voids they find differ greatly, finding median effective radii of 24 and 16 \h Mpc respectively. 
Differences in smoothing scale, spatial resolution, sampling density, and survey boundary conditions further contribute to variations in the inferred void locations and extents. In addition, cosmic voids possess highly irregular, non-spherical, and hierarchical morphologies, resulting in different identification schemes naturally partitioning the same under-dense volume differently.

\section{Conclusions}\label{sec:conclusion}

We have presented catalogues of dynamically defined cosmic voids and knots identified using the V-web formalism and the CF4++ZOA reconstructed density and velocity fields. Unlike traditional density-based approaches, the V-web classification is derived directly from the eigenvalues of the velocity shear tensor which traces the underlying gravitational flow field. Providing a physically motivated description of the large-scale structure, which highlights regions of coherent expansion and convergence that correspond to cosmic voids and knots, respectively.

Applying our procedure to CF4++ZOA, we detect 37 void and 42 knot regions within the surveyed volume, which are distinct and robust. Within the CF4++ galaxy sample, 3,041 are located within our identified void regions while 7,346 reside within knot regions. In contrast, a simple density-based classification yields 23,020 galaxies in under-dense regions ($\delta<0$) and 42,313 in over-dense regions. The catalogues  presented in this letter are constructed using a hierarchical threshold approach combined with a flood-fill algorithm, each of which were validated using the ensemble of HMC realizations. By requiring structures to be recovered in at least 68\% of the realizations and by excluding regions significantly affected by survey boundaries, we obtain a catalogue of dynamically significant structures that is resilient to reconstruction uncertainties.

The resulting void and knot distribution reproduces the main features of the local cosmic web traced by the peculiar velocity field. Prominent structures—including the Local Void, Laniakea, Shapley, Hercules, Perseus, Pisces, Vela, and Columba–Lepus regions—are recovered as distinct dynamical features. Comparisons with recent void catalogues show broad agreement for the largest underdense regions, while underscoring the strong methodological dependence in void identification. 

The catalogues presented here provide a reference framework for studies of the local Universe based on peculiar velocity reconstructions. They can be used to investigate environmental influences on galaxy evolution, characterize large-scale gravitational flows, and provide targets for future comparisons with numerical simulations and next-generation surveys. More broadly, this work demonstrates the value of velocity-field reconstructions for mapping the cosmic web and identifying its most prominent expanding and converging structures in a dynamically consistent manner.
The V-web classifications and accompanying catalogues presented in this work will be released publicly via GitHub upon publication. 
\begin{acknowledgements}
HMC acknowledges support from the Institut Universitaire de France and from Centre National d’Etudes Spatiales (CNES), France. 
\end{acknowledgements}

\bibliographystyle{aa} 
\bibliography{biblio} 

\begin{appendix}

\section{Dependence on the V-web Threshold Parameter}

The choice of $\lambda_{th}$ provides a direct dynamical interpretation of the V-web classification, whereby regions are distinguished solely by the sign of the velocity-shear tensor eigenvalues. Throughout this work, voids and knots are identified using the criteria described in Section \ref{sec:methodology}, adopting a threshold of $\lambda_{th}=0$. We ultimately adopt this value to remain consistent with previous CosmicFlows analyses. However, because the number and extent of identified structures depend on the adopted threshold, it is useful to assess the sensitivity of our catalogue to this choice.

More recent studies have proposed scale- and redshift-dependent values of $\lambda_{th}$ \citep[e.g.][]{olex_universal_2025}, calibrated using N-body simulations, with the aim of providing a physically motivated threshold rather than adopting a single fixed value throughout the volume. Implementing a comparable calibration for the CF4++ZOA reconstruction would require a detailed analysis of mock reconstructions and associated simulations, which lies beyond the scope of the present work. We therefore restrict ourselves to examining another commonly used fixed threshold within the V-web formalism, $\lambda_{th}=0.44$ \citep[e.g.][]{hoffman_kinematic_2012,libeskind_tracing_2018}.

\begin{figure*}[hbt!]
    \centering
    \includegraphics[width=1\linewidth]{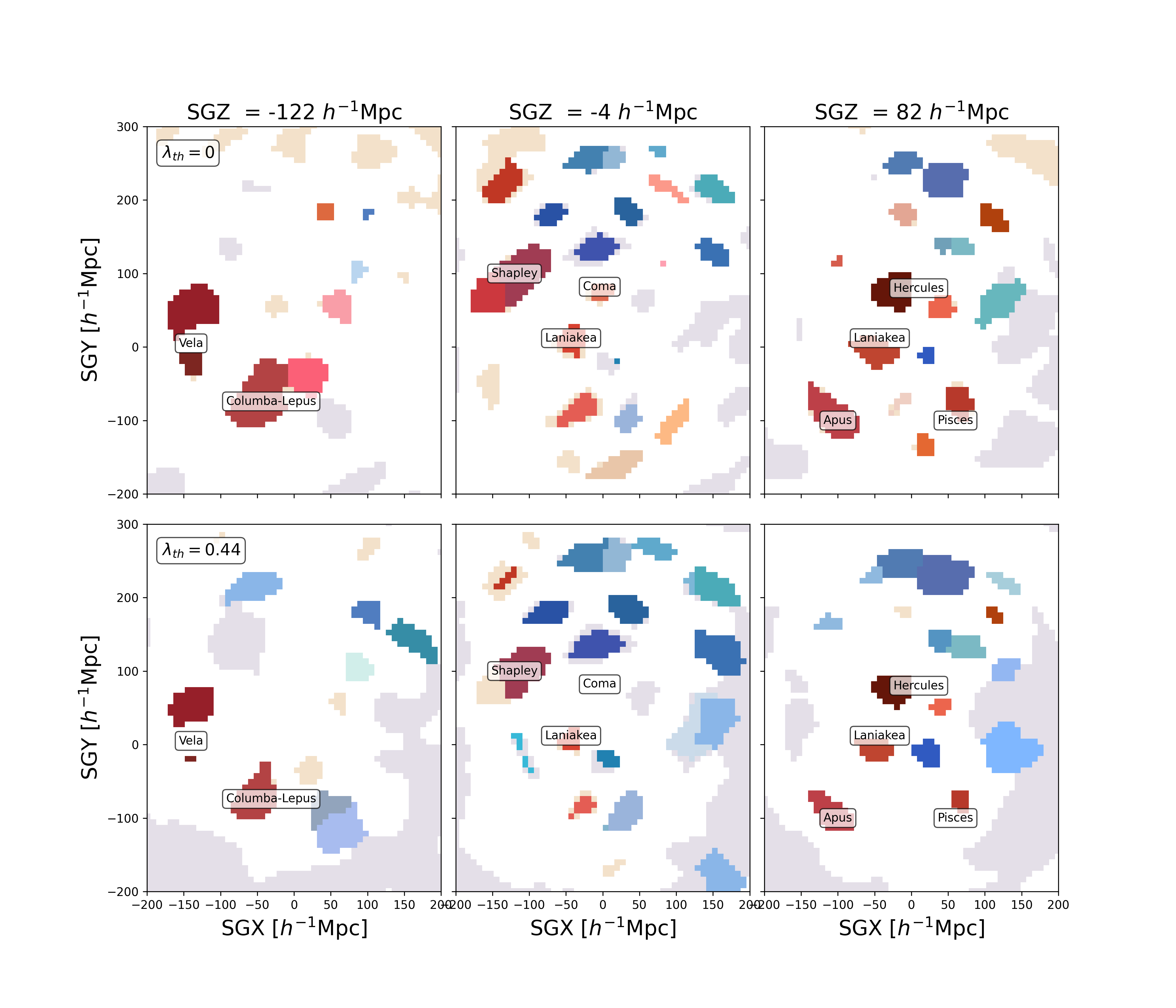}
    \caption{The coloured regions depict the spatial distribution of the  structure groupings that were identified in the full CF4++ZOA reconstruction assuming $\lambda_{\rm th}=0$ (top) and  $\lambda_{\rm th}=0.44$ (bottom), through various slices of SGZ. The light lilac and orange regions, respectively, show all the void and knot regions which are classified for the full reconstruction. 
    The dark blue-toned (void) and red-toned (knot) regions indicate structures that satisfy the selection criteria described in Section~\ref{sec:methodology}, the regions in the top panels additionally correspond to the values in Tables \ref{tab:vv} and \ref{tab:structures}.}
    \label{fig:app}
\end{figure*}

The coloured regions of Figure \ref{fig:app} show the spatial distribution of the structure groupings identified within the full CF4++ZOA reconstruction for $\lambda_{th}=0$  (top panels) and $\lambda_{th}=0.44$  (bottom panels), through various slices of SGZ. The light lilac and orange shading denote all regions classified as voids and knots, respectively, in the full reconstruction, while the Dark blue (void) dark red (knot) shading highlight the subset of structures that satisfy the selection criteria described in Section \ref{sec:methodology}. The selected structures in the top panels correspond to the values in Tables \ref{tab:vv} and \ref{tab:structures}. Applying the same methodology with $\lambda_{th}=0.44$  produces substantially different final catalogues of structures. In this case, 61 robust voids and 22 knots are identified, compared with the 37 voids and 42 knots presented in the main text. 

The higher threshold increases the number of identified voids while reducing the number of knots, reflecting the more stringent collapse criterion imposed by a larger value of $\lambda_{th}$ . This difference is evident in the lower panels of the figure, where changes in the extent and connectivity of individual structures can be seen relative to the $\lambda_{th}=0$ case. Despite these differences, the most prominent voids and over-dense regions remain readily identifiable under both threshold choices, demonstrating that the principal features of the reconstructed cosmic web are robust. The primary effect of increasing $\lambda_{th}$  is therefore to modify the boundaries, connectivity, and classification of marginal regions, rather than to alter the existence of the largest structures themselves. These comparisons highlight the sensitivity of V-web catalogues to the adopted threshold parameter while also illustrating the stability of the dominant large-scale features recovered by the V-web formalism.

\section{Full Catalogues}\label{app:A}

In this section, we expand the discussion presented in the main text to the full catalogue. The top panel of Figure \ref{fig:app} shows the distinct void and knot regions sliced through the various SGZ planes. The left panel is centred on SGZ=-122 \h Mpc plane, where Vela and Columba-Lepus are prominent; the middle  shows the SGZ = 4 \h Mpc plane, featuring Coma, Laniakea and Shapley; while the right panel is sliced through SGZ = 82 \h Mpc displaying Hercules, Apus, Laniakea and Pisces.
 We also present the extended versions of the void and knot catalogues, of Table \ref{tab:voids}. 
 
 For Table \ref{tab:vv}: Columns 1 and 2 have the ID of the void and the minimum $\lambda_{th}$ that would be required for the central voxel to be classified as a void. Columns (3-5) represent the coordinates of the centre in terms of equatorial coordinates and comoving distance. Column 6 is the constellation region within which the centre is located. Column 7 is the centre position in supergalactic coordinates. Columns 8-9 are respectively the full volume of the void and its effective radius, while the bracketed values are the volume and radius corresponding to our $1\sigma$ requirement for the HMC realizations.
 
 For table \ref{tab:structures}: Column 1 is the ID of the knot, Column 2 lists the supercluster the knot would be associated with in \cite{hollinger_hidden_2026}; Column 3 is the  maximum $\lambda_{th}$ that could be used to classify the region's central voxel  as a knot; Columns 4-8 are the equatorial coordinates, comoving distance and volume; finally, Column 9 lists the mass of the knot region, with the values in brackets having the same meaning as before.

\begin{table*}
\centering
\begin{tabular}{ccccccccc}
\hline
Void &  Detected & R.A & Dec. & $D_c$  & Constellation & SGX,SGY,SGZ & Volume & Effective Radius   \\
ID &  $\lambda$ &[$^\circ$]& [$^\circ$] & [\h Mpc]   & Region &  [\h Mpc] & [$10^3 h^{-3}$Mpc$^3$]  &  [\h Mpc] \\
\hline
1 & -3.34 & 197.6 & 16.7 & 141 & Coma Berenices & (-27,137,20) & 155.0 (125.9) & 33.3 (31.1) \\
2 & -2.18 & 204.5 & 39.8 & 239 & Canes Venatici & (43,223,74) & 156.4 (108.2) & 33.4 (29.6) \\
3 & -1.83 & 191.4 & 64.6 & 86 & Draco & (51,66,20) & 43.4 (31.5) & 21.8 (19.6) \\
4 & -1.53 & 180.7 & 35.3 & 187 & Ursa Major & (35,184,-4) & 27.7 (15.7) & 18.8 (15.5) \\
5 & -1.51 & 195.3 & 7.2 & 196 & Virgo & (-66,184,12) & 48.2 (23.4) & 22.6 (17.7) \\
6 & -1.43 & 207.9 & 27.2 & 270 & Boötes & (-12,254,90) & 48.6 (20.5) & 22.6 (17.0) \\
7 & -1.41 & 129.3 & 50.2 & 236 & Ursa Major & (160,152,-82) & 164.0 (71.5) & 34.0 (25.8) \\
8 & -1.4 & 253.0 & 49.1 & 246 & Hercules & (90,137,184) & 100.6 (61.5) & 28.9 (24.5) \\
9 & -1.32 & 243.7 & 45.3 & 236 & Hercules & (66,152,168) & 102.5 (69.6) & 29.0 (25.5) \\
10 & -1.28 & 296.7 & 22.9 & 93 & Vulpecula & (20,-12,90) & 49.1 (17.2) & 22.7 (16.0) \\
11 & -1.24 & 68.9 & -42.9 & 46 & Caelum & (-12,-27,-35) & 28.6 (20.5) & 19.0 (17.0) \\
12 & -1.19 & 137.3 & 61.8 & 196 & Ursa Major & (145,129,-27) & 140.7 (68.2) & 32.3 (25.3) \\
13 & -1.18 & 328.4 & -16.8 & 34 & Capricornus & (-4,-27,20) & 35.3 (28.1) & 20.3 (18.9) \\
14 & -1.13 & 153.4 & 33.8 & 202 & Leo Minor & (66,176,-74) & 173.1 (76.8) & 34.6 (26.4) \\
15 & -1.06 & 243.2 & 7.5 & 252 & Hercules & (-90,137,191) & 95.4 (17.6) & 28.3 (16.1) \\
16 & -1.05 & 161.3 & 21.7 & 216 & Leo & (20,199,-82) & 62.0 (23.8) & 24.6 (17.9) \\
17 & -1.03 & 148.8 & -32.4 & 114 & Antlia & (-66,43,-82) & 78.7 (27.2) & 26.6 (18.7) \\
18 & -1.02 & 186.6 & 20.9 & 255 & Coma Berenices & (-20,254,-4) & 52.9 (20.0) & 23.3 (16.8) \\
19 & -1.01 & 159.3 & 26.3 & 244 & Leo Minor & (43,223,-90) & 34.8 (12.4) & 20.3 (14.4) \\
20 & -0.99 & 176.1 & 37.0 & 278 & Ursa Major & (66,270,-20) & 46.7 (11.9) & 22.3 (14.2) \\
21 & -0.97 & 224.6 & 41.8 & 178 & Boötes & (35,145,98) & 28.6 (14.8) & 19.0 (15.2) \\
22 & -0.97 & 247.8 & 14.7 & 52 & Hercules & (-12,27,43) & 15.7 (10.0) & 15.5 (13.4) \\
23 & -0.95 & 159.1 & 56.8 & 269 & Ursa Major & (160,215,-20) & 64.8 (11.0) & 24.9 (13.8) \\
24 & -0.93 & 271.3 & -58.8 & 107 & Pavo & (-98,-35,27) & 91.6 (38.6) & 28.0 (21.0) \\
25 & -0.87 & 237.6 & 33.8 & 161 & Corona Borealis & (12,113,113) & 38.1 (17.6) & 20.9 (16.1) \\
26 & -0.86 & 174.6 & 11.5 & 168 & Leo & (-27,160,-43) & 19.6 (8.6) & 16.7 (12.7) \\
27 & -0.79 & 184.2 & 27.6 & 254 & Coma Berenices & (12,254,-4) & 24.8 (6.2) & 18.1 (11.4) \\
28 & -0.78 & 9.9 & -5.8 & 104 & Cetus & (35,-98,4) & 31.5 (8.1) & 19.6 (12.5) \\
29 & -0.78 & 32.7 & -28.0 & 107 & Fornax & (4,-98,-43) & 32.9 (5.2) & 19.9 (10.8) \\
30 & -0.75 & 212.0 & 56.2 & 169 & Ursa Major & (74,137,66) & 29.1 (10.5) & 19.1 (13.6) \\
31 & -0.71 & 284.9 & 65.6 & 176 & Draco & (121,59,113) & 231.7 (26.2) & 38.1 (18.4) \\
32 & -0.67 & 136.7 & 39.6 & 113 & Lynx & (59,82,-51) & 20.0 (8.6) & 16.8 (12.7) \\
33 & -0.61 & 175.5 & 58.7 & 260 & Ursa Major & (145,215,20) & 78.7 (1.0) & 26.6 (6.1) \\
34 & -0.54 & 153.5 & -36.6 & 77 & Antlia & (-51,27,-51) & 27.7 (5.7) & 18.8 (11.1) \\
35 & -0.51 & 173.2 & 22.2 & 204 & Leo & (4,199,-43) & 10.5 (1.9) & 13.6 (7.7) \\
36 & -0.51 & 116.5 & 22.1 & 198 & Gemini & (90,90,-152) & 54.8 (2.4) & 23.6 (8.3) \\
37 & -0.41 & 174.9 & -18.4 & 72 & Crater & (-43,51,-27) & 13.4 (1.9) & 14.7 (7.7) \\

\hline
\end{tabular}
\caption{Catalogue of voids: (1) Void IDs, based on earliest detection in the mean CF4++ZOA reconstruction; (2) Value of $\lambda_{th}$ where minimum is first detected; (3) The constellation in which the centre is located; (4-6) equatorial coordinates and distance of the minima; (7) Supergalactic Cartesian position; (8) total volume : based on the mean CF4++ZOA  reconstruction (identified in 68\% of the HMC realizations); (9) Effective radii.}
\label{tab:vv}
\end{table*}

\begin{table*}
\centering
\begin{tabular}{ccccccccc}
\hline
Knot & Corresponding &  Detected & R.A & Dec. & $D_c$  &  SGX,SGY,SGZ & Volume & Mass  \\
ID & Supercluster & $\lambda$ &[$^\circ$]& [$^\circ$] & [\h Mpc]   &  [\h Mpc] & [$10^3 h^{-3}$Mpc$^3$]  &  [$\times10^{15}$M$_{\odot}$] \\
\hline
1 & Hercules  & 3.98 & 238.0 & 15.5 & 114 & (-27,74,82) & 107.8(90.6) & 38.9(33.9)  \\
2 & Shapley & 2.7 & 199.8 & -15.5 & 155 & (-105,113,4) & 171.7(133.0) & 59.0(48.1)  \\
3 & Vela & 2.32 & 123.1 & -49.4 & 201 & (-121,-12,-160) & 145.4(88.2) & 44.2(29.2)  \\
4 & Vela* & 2.12 & 153.7 & -39.6 & 186 & (-129,59,-121) & 142.6(86.3) & 42.5(28.1)  \\
5 & Columba-Lepus  & 2.09 & 80.4 & -48.5 & 154 & (-59,-74,-121) & 322.3(187.9) & 85.5(53.5)  \\
6 & --  & 2.09 & 220.2 & 55.9 & 227 & (98,176,105) & 84.9(66.8) & 27.1(22.2)  \\
7 & Laniakea  & 2.08 & 273.9 & -10.3 & 93 & (-43,-4,82) & 105.9(73.0) & 27.9(20.6)  \\
8 & Pisces & 2.07 & 350.3 & 18.5 & 130 & (74,-90,59) & 121.1(75.8) & 32.1(21.4)  \\
9 & Apus & 1.92 & 305.5 & -46.5 & 158 & (-105,-98,66) & 330.0(166.9) & 87.6(49.5)  \\
10 & Perseus & 1.74 & 55.2 & 11.1 & 102 & (66,-51,-59) & 129.7(79.6) & 33.2(21.8)  \\
11 & Pisces* & 1.57 & 339.0 & 50.1 & 63 & (51,-12,35) & 50.5(37.7) & 14.3(10.9)  \\
12 & Shapley* & 1.57 & 205.4 & -39.6 & 173 & (-160,66,-4) & 98.7(65.3) & 29.3(20.5)  \\
13 & --  & 1.35 & 196.6 & -0.5 & 261 & (-121,230,12) & 226.0(70.6) & 53.6(19.8)  \\
14 & Laniakea*  & 1.33 & 224.6 & -53.4 & 51 & (-51,4,4) & 48.2(35.3) & 11.7(8.9)  \\
15 & --  & 1.28 & 252.2 & 15.1 & 177 & (-35,82,152) & 37.2(18.1) & 8.8(4.6)  \\
16 & --  & 1.2 & 179.4 & 9.4 & 225 & (-51,215,-43) & 108.7(60.1) & 30.2(18.2)  \\
17 & Perseus & 1.2 & 61.5 & 44.7 & 62 & (59,-4,-20) & 77.2(54.8) & 19.7(14.6)  \\
18 & --  & 1.15 & 351.9 & -35.0 & 85 & (-20,-82,12) & 88.7(34.3) & 20.3(9.1)  \\
19 & --  & 1.14 & 271.4 & 66.3 & 268 & (176,113,168) & 139.2(58.2) & 36.0(16.9)  \\
20 & Hercules* & 1.14 & 260.9 & 50.1 & 106 & (43,51,82) & 36.7(20.5) & 9.8(5.8)  \\
21 & Shapley* & 1.08 & 219.5 & -6.9 & 174 & (-105,121,66) & 21.9(8.6) & 5.7(2.4)  \\
22 & --  & 1.02 & 332.9 & -1.1 & 169 & (27,-129,105) & 71.0(14.8) & 13.5(3.1)  \\
23 & Coma  & 0.93 & 161.4 & 30.2 & 89 & (20,82,-27) & 30.5(19.6) & 8.3(5.5)  \\
24 & --  & 0.9 & 146.3 & 18.1 & 232 & (35,184,-137) & 32.4(12.4) & 6.5(2.7)  \\
25 & --  & 0.88 & 155.2 & -4.6 & 132 & (-35,98,-82) & 42.9(17.6) & 8.3(3.7)  \\
26 & Columba-Lepus*  & 0.85 & 84.8 & -14.1 & 128 & (20,-35,-121) & 82.5(23.4) & 20.6(6.3)  \\
27 & --  & 0.83 & 160.6 & 27.5 & 148 & (27,137,-51) & 17.2(8.6) & 3.7(1.9)  \\
28 & --  & 0.78 & 181.9 & 53.9 & 191 & (90,168,20) & 11.9(5.7) & 2.8(1.4)  \\
29 & --  & 0.77 & 250.4 & 52.2 & 182 & (74,105,129) & 19.6(6.7) & 4.6(1.7)  \\
30 & --  & 0.76 & 114.6 & 17.0 & 147 & (59,59,-121) & 42.4(4.3) & 7.5(0.9)  \\
31 & --  & 0.75 & 210.6 & 28.8 & 198 & (-4,184,74) & 14.8(4.3) & 3.2(1.0)  \\
32 & --  & 0.71 & 171.8 & 42.5 & 238 & (82,223,-20) & 42.9(8.1) & 9.3(1.8)  \\
33 & --  & 0.69 & 26.3 & -11.5 & 166 & (51,-152,-43) & 93.0(23.4) & 21.2(6.0)  \\
34 & --  & 0.68 & 22.4 & 12.1 & 139 & (90,-105,-12) & 69.1(1.4) & 15.2(0.3)  \\
35 & --  & 0.66 & 11.5 & -22.7 & 169 & (12,-168,-12) & 81.5(17.2) & 17.0(4.2)  \\
36 & --  & 0.65 & 125.0 & 23.7 & 280 & (113,160,-199) & 73.0(19.6) & 18.5(5.7)  \\
37 & --  & 0.64 & 148.2 & 52.7 & 136 & (82,105,-27) & 22.9(7.6) & 5.2(1.8)  \\
38 & --  & 0.63 & 325.4 & -14.9 & 106 & (-12,-82,66) & 28.6(2.4) & 5.3(0.5)  \\
39 & --  & 0.6 & 174.7 & -79.7 & 133 & (-121,-35,-43) & 108.7(2.4) & 23.9(0.6)  \\
40 & --  & 0.52 & 232.9 & 8.6 & 225 & (-82,152,145) & 15.3(1.9) & 3.5(0.5)  \\
41 & --  & 0.46 & 180.9 & -5.8 & 243 & (-113,207,-59) & 17.6(1.0) & 3.9(0.2)  \\
42 & --  & 0.28 & 230.4 & 31.3 & 217 & (4,168,137) & 14.3(0.5) & 2.6(0.1)  \\

\hline
\end{tabular}
\caption{Catalogue of knot regions: (1) Knot IDs, based on earliest detection in the mean CF4++ZOA reconstruction; 
(1) Name of corresponding supercluster from \cite{hollinger_hidden_2026};
(2) Value of $\lambda_{th}$ where maxima is first detected; 
(4-6) equatorial coordinates and distance of the maxima; (7) Supergalactic Cartesian position; (8-9) Volume and mass corresponding to the mean CF4++ZOA  reconstruction (identified in 68\% of the HMC realizations).}
\label{tab:structures}
\end{table*}

\end{appendix}

\end{document}